body of paper

\documentstyle[preprint,eqsecnum,aps]{revtex}
\begin{document}
\draft
\title{
Long wavelength iteration of Einstein's equations near a
spacetime singularity}
\author{Nathalie Deruelle}
\address{D\'epartement d'Astrophysique Relativiste et de Cosmologie,\\
Centre National de la Recherche Scientifique, \\Observatoire de
Paris, 92195 Meudon, France\\
and Department of Applied Mathematics and Theoretical Physics,\\
University of Cambridge, Silver Street,\\ Cambridge CB3 9EW,
England}
\author{David Langlois}
\address{The Racah Institute of Physics, The Hebrew University,\\
Givat Ram, Jerusalem 91904, Israel \\
 and  D\'epartement d'Astrophysique Relativiste et de Cosmologie,\\
Centre National de la Recherche Scientifique,\\ Observatoire de
Paris, 92195 Meudon, France}
\date{\today}
\maketitle
\begin{abstract}
We clarify the links between a recently developped long wavelength iteration
scheme of
Einstein's equations, the  Belinski Khalatnikov Lifchitz (BKL) general
solution near a singularity and
the antinewtonian scheme of Tomita's. We determine the regimes when
the long wavelength or antinewtonian scheme is directly applicable and show how
it can otherwise be
implemented to yield the BKL oscillatory approach to a spacetime
singularity. When directly applicable we obtain the generic solution
of the scheme at first iteration (third order in the gradients) for
matter a perfect fluid. Specializing to
spherical symmetry for simplicity and to clarify gauge issues, we then show
how the metric behaves near a singularity when gradient effects are
taken into account.
\end{abstract}


\section{Introduction}

Despite the fact that the universe is clearly inhomogeneous on galactic
scale and the possiblity, raised by some inflationary scenarios (see
e.g.) \cite{linde}, that its geometry be ``chaotic" on scales larger
than the Hubble radius, most cosmological models are still the
homogeneous and isotropic model of Friedman, Robertson and Walker (FRW).
Many convincing reasons, physical or philosophical, can be given to that
state of affair, but there is also a purely technical one~:  very few
inhomogeneous solutions of cosmological interest are known (see, e.g.
\cite{krasinski}).

Various approximate solutions however have been given in the past.  A
simple one is the ``quasi-isotropic" solution of Lifchitz and
Khalatnikov (see e.g.  \cite{ll}) the spatial sections of which (in a
synchronous reference frame) are just uniformly stretched in the course
of time ($ds^2=-dt^2+a^2(t)h_{ij}dx^idx^k$, where the arbitrary``seed"
metric $h_{ij}(x^k)$ depends on space only).  This metric is exact and
reduces to the standard FRW metric if the spatial sections are maximally
symmetric, and is a good approximation to an exact solution of
Einstein's equations if, as we shall recall below, all spatial
derivatives remain small, that is if all ``point to point" interactions
are neglected.

A more general approximate solution when all gradients are neglected is
the ``antinewtonian" solution of Tomita's \cite{tomita}, which, as we
shall recall, depends on as many arbitrary functions as a generic
solution of Einstein's equations.

Finally the ``general oscillatory solution" studied by Belinski,
Lifchitz and Khalatnikov \cite{bkl} is the most elaborate approximate
description of a generic solution of Einstein's equations near the
Big-Bang.

There was recently a renewed interest in these approximation solutions
first of all because the observations of the COBE satellite urged a
fresh view on the old problem of structure formation (see e.g.
\cite{lily}) but also because a new line of attack of Einstein's
equations was pursued.  Indeed, in a series of papers \cite{ss}, Salopek
Stewart and collaborators developped a ``long-wavelength" iteration
scheme not of Einstein's equations but of the Hamilton-Jacobi equation
for General Relativity.  Their method, which consists at lowest order in
neglecting all spatial gradients, leads back in most instances to the
quasi-isotropic solution mentionned above, and yielded, for dust at
least, the solution up to and including the third iteration (that is
accurate to order 6 in the gradients as will become clear below
\cite{parry}).  (The case of a more general perfect fluid is more
awkward to handle in this Hamiltonian formalism.)

In another series of papers the present authors (together with Comer,
Goldwirth, Tomita and Parry) \cite{cdlp}, \cite{dg}, \cite{td}
iterated in the same way the Einstein equations themselves.  They noted
that the zeroth order quasi-isotropic solution, although not generic, is
an attractor at late cosmological times of the generic solution of
Tomita.  Concentrating then on this quasi-isotropic limit of the zeroth
order solution they obtained the solution up to and including the second
iteration (5th order in the gradients) for matter being a perfect fluid
with constant adiabatic index or a scalar field.  In the particular case
of dust their result is identical to that of \cite{ss} so that the link
between the two methods could be clearly made.

The motivation for going beyond the zeroth order is the wish to describe
inhomogeneities within the Hubble radius.  Indeed the approximation at
the root of these long wavelength iteration schemes is the following.
Take a synchronous reference frame where the line element reads:
$$ds^2=-dt^2+\gamma_{ij}(x^k,t)dx^idx^k\qquad (i,j,=1,2,3).$$
At each point define a local scale factor $a$ and a Hubble time $H^{-1}$ by
$$a^2\equiv ({\rm det}\gamma_{ij})^{1/3}\quad,\quad H\equiv\dot a/a,$$
where $\dot a\equiv\partial a/\partial t$.  The Hubble time is the
characteristic proper time on which the metric evolves.  The
characteristic comoving length on which it varies is denoted $L$ :
$\partial_i\gamma_{jk}\approx L^{-1}\gamma_{ij}$.  The long wavelength
approximation is the assumption that the characteristic scale of spatial
variation is much bigger than the Hubble radius, that is:
$${1\over a}\partial_i\gamma_{jk} \ll\dot\gamma_{ij}\quad
\Longleftrightarrow\quad aL\gg H^{-1}.$$
At lowest order then the long wavelength approximation is not suited to
describe e.g. the formation of structure within the present Hubble radius.
One can hope however that the iteration scheme pushed at a sufficiently high
order can give results valid within the Hubble radius. Some numerical
investigation of this question has been undertaken by Deruelle and Goldwirth
\cite{dg} (see also \cite{ss}) but further work is nevertheless required to
assess the convergence properties of the approximation scheme.

Another motivation to go beyond lowest order, which is the one for this paper,
is to study, within a long wavelength approximation scheme, the behaviour of a
generic solution of Einstein's equations near a space-time singularity and make
the link between that scheme and the BKL general solution referred to above.

The point of view is therefore very different from the one adopted in
our previous papers since, instead of the late time quasi-isotropic
solution, we consider here the solution near a singularity, thus going
backward in time.  In this case the quasi-isotropic behaviour is no
longer valid and the approximate solution becomes more complicated since
it is no longer possible to separate the time dependence and the spatial
dependence into a scale factor and a ``seed" metric respectively.  The
generic (i.e.  without assuming quasi-isotropy) first order solution was
given by Tomita in the case of dust and radiation (it can also be found
in \cite{ss}).  Here we consider the more general case of a barytropic
perfect fluid with an equation of state of the form $p/\epsilon=\Gamma-1$,
where $\Gamma$ is a constant.  Although an explicit solution for the first
order solution cannot be given for $\Gamma$ different from $\Gamma=1$ (dust)
and $\Gamma=2$ (stiff matter), an explicit limit near the singularity
can be given in all cases (sections 2 and 3).

Once the first order solution has been given, we analyze the validity of
the approximation near the singularity (section 4).  To do this we
examine the time evolution of the terms which were neglected at first
order.  We find that they could not always be ignored and we give a
condition of validity for the approximation scheme.  In the cases when
this condition is not fulfilled, we are able to make the link with the
work of Belinski Kalatnikov and Lifschitz.  We believe that the way we
recover the oscillatory behaviour of the metric, which does not
introduce intermediate Bianchi IX geometries, is more straightforward
than the original approach of BKL, and will allow in particular an
easier analysis of the genericity of the ``spindle" singularities found
by Bruni et al \cite{bruni}.

We then give the generic third order solution (section 5). For the sake of
simplicity we then apply in detail the approximation scheme to the case of
spherical symmetry (section 6).  In particular we show that, in the case of
dust, the third order solution corresponds to an expansion of the Tolman-Bondi
solution in time. Finally, in section 7, we give our conclusions and comment
on the usefulness of the long wavelentgh approximation.

\section{The long wavelength iteration scheme}

In this section we first rewrite Einstein's equations in a way
convenient for our purposes and then describe the iteration procedure.

We place ourselves in a synchronous reference frame where the line element
takes the form:
\begin{equation}
ds^2=-dt^2+\gamma_{ij}(t,x^k)dx^idx^j\qquad (i,j=1,2,3)
\end{equation}
(Coordinates transformations involving four functions of space can
still be performed without spoiling the synchronicity of the
reference frame; see e.g. \cite{ll}.) Matter is taken to be a perfect fluid
with pressure $p$, energy density $\epsilon$, unit four velocity $u^\mu$
($\mu=0,1,2,3$) and stress energy tensor:
\begin{equation}
T_{\mu\nu}=(\epsilon+p)u_\mu u_\nu+pg_{\mu\nu}
\label{setensor}\end{equation}
with the further restriction that $p/\epsilon=\Gamma-1$ where the
 index $\Gamma$ is supposed to be constant, positive and less
than to 2 (the limiting cases $\Gamma=0$ and $\Gamma=2$ correspond
respectively to  a cosmological constant and a ``stiff'' fluid whose
speed of sound equals the speed of light; $\Gamma=1$ is dust,
$\Gamma=4/3$ radiation; fluids with $0<\Gamma<2/3$ violate the strong
energy condition and can be called ``inflationary''.)

Einstein's equations are : $R^\mu_\nu=S^\mu_\nu$ with
$S^\mu_\nu\equiv\chi (T^\mu_\nu-{1\over2}\delta^\mu_\nu T^\rho_\rho)$
and $\chi\equiv 8\pi G$, $G$ being Newton's constant. In a synchronous
reference frame the components of the Ricci tensor $R^\mu_\nu$ are (see e.g.
\cite{ll}):
\begin{equation}
R^0_0={\scriptstyle{1\over2}}\dot\kappa
+{\scriptstyle{1\over 4}}\kappa^i_j\kappa^j_i \quad
{,}\quad
R^0_i=-{\scriptstyle{1\over 2}}(\kappa^j_{i;j}-\kappa_{,i})\quad{,}\quad
R^i_j=\tilde R^i_j+{\scriptstyle{1\over 2}}\dot\kappa^i_j+{\scriptstyle{1\over
 4}}\kappa\kappa^i_j
\end{equation}
where $\kappa_{ij}\equiv\dot\gamma_{ij}$ is the extrinsic curvature (a
dot denotes the derivative with respect to time $t$, a semicolon the
covariant derivative with respect to $\gamma_{ij}$); all indices are
raised with the inverse metric $\gamma^{ij}$; $\kappa\equiv\kappa^i_i$
and $\tilde R^i_j$ is the Ricci tensor associated with the metric
$\gamma_{ij}$.

Now, one can always decompose $\kappa^i_j\equiv\gamma^{ik}\dot\gamma_{kj}$
into a trace and traceless part:
\begin{equation}
\kappa^i_j=2H\delta^i_j +A^i_j/a^3 \label{extr}
\end{equation}
where the ``anisotropy matrix'' $A^i_j$ is traceless ($A^i_i=0$) and
where we have introduced a local ``scale factor'' $a\equiv ({\rm
det}\gamma_{ij})^{1/6}$, so that the
local ``Hubble parameter'' $H$ is $H\equiv\dot a/a$. When $a(t,x^k)$ and
$A^i_j(t,x^k)$ are known the metric $\gamma_{ij}$ is obtained by
integrating the 6 linear equations:
\begin{equation}
\dot\gamma_{ij}=2H\gamma_{ij}+A^k_j\gamma_{ki}/a^3.\label{e5}
\end{equation}
(The matrix $A^i_j$ is therefore such that
$A^k_i\gamma_{kj}=A^k_j\gamma_{ki}$.)

Let us then rewrite Einstein's equations as equations for $a$ and
$A^i_j$.
The traceless part of $R^i_j=S^i_j$ gives:
\begin{equation}
\dot A^i_j=2a^3(\bar S^i_j-\bar R^i_j) \label{einstein1}
\end{equation}
where $\bar R^i_j\equiv\tilde R^i_j-{1\over3}\delta^i_j\tilde R$ and $\bar
S^i_j\equiv \chi\epsilon\Gamma(u^iu_j-{1\over3}\delta^i_ju^ku_k)$ are
the traceless parts of $\tilde R^i_j$ and $S^i_j$.

The trace of $R^i_j=S^i_j$ together with the $(^0_0)$ equation
$R^0_0=S^0_0$ give:
\begin{equation}
2\dot H+3\Gamma H^2+{||A||^2\over8 a^6}(2-\Gamma)=
{\scriptstyle{1\over6}}(2-3\Gamma)\tilde R
-{\scriptstyle{1\over 3}}\chi\epsilon(4-3\Gamma)\Gamma u^ku_k \label{einstein2}
\end{equation}
\begin{equation}
\chi\epsilon(1+\Gamma u^ku_k)=3 H^2-{||A||^2\over8a^6}+{\scriptstyle{1\over 2}}
\tilde R
\label{einstein3}
\end{equation}
where $||A||^2\equiv A^i_jA^j_i$.

Finally the $(^0_i)$ equation $R^0_i=S^0_i$ reads:
\begin{equation}
2\partial_i H-{1\over2}\left({A^j_i\over a^3}\right)_{;j}
=\chi\epsilon\Gamma u_i\sqrt{1+u^ku_k}. \label{einstein4}
\end{equation}

Equations (\ref{einstein1}-\ref{einstein4}) are strictly equivalent
 to Einstein's equations but
are written in a form suitable for the implementation of the long
wavelength iteration scheme. We shall also use the following
consequence of Einstein's equations (obtained by differentiating
(\ref{einstein3})
and using (\ref{einstein1}-\ref{einstein4})):
\begin{equation} (1-\Gamma)\partial_i\epsilon=\Gamma\left\{D_j(\epsilon
u^ju_i)+{1\over a^3}\left[\epsilon u_ia^3
\sqrt{1+u^ku_k}\right]^.\right\}.\label{e10}
\end{equation}

The  long wavelength approximation consists in neglecting all the terms
quadratic in the gradients, that is in the spatial derivatives, in
Einstein's equations.  Now, from Eq (\ref{einstein4}) or (\ref{e10}) the
3-velocity $u_i$ is at least first order in the gradients so that the
right-hand side of Eq (\ref{einstein1}) is at least second order.  At
first order then it can be set equal to zero so that Eq (\ref{einstein1})
trivially gives that the anisotropy matrix $A^i_j$ does not depend on
time:  $A^i_j\simeq {}^{(1)}A^i_j(x^k)$.  Then Eq (\ref{einstein2}), the
right hand side of which can be ignored at first order, is an equation
which, when integrated, gives the scale factor $^{(1)}a(t,x^k)$.  The
anisotropy matrix and the scale factor being known, Eq (\ref{e5}) yields
the first order metric $^{(1)}\gamma_{ij}(t,x^k)$.  Finally Eq
(\ref{einstein3}- \ref{einstein4}) where at first order $\tilde R$ and
$u^ku_k$ can be ignored give the energy density $^{(1)}\epsilon(t,x^k)$
of the fluid as well as its 3-velocity $^{(1)}u_i(t,x^k)$ (the three
velocity can equivalently be obtained from (\ref{e10})).  This first
order solution is given by Eardley, Liang and Sachs \cite{eardley}, and
Tomita \cite{tomita}.  It is reviewed in the next Section.

\section{The generic first order solution}

In this section we give the general solution  (see \cite{eardley} and
\cite{tomita}) of the truncated Einstein equations (\ref{einstein1}-
\ref{einstein4}) in which all terms of order greater than one in the
spatial derivatives are neglected.

\subsection{The anisotropy matrix}

As already mentioned in section 2, Eq (\ref{einstein1}) at lowest order
reads $\dot A^i_j=0$ and gives that the anisotropy matrix depends on
space alone:
\begin{equation}
A^i_j ={}^{(1)}A^i_j(x^k)\quad\hbox{with}\quad ^{(1)}A^i_i=0.
\label{e11}\end{equation}

\subsection{The scale factor}

As for the Eq (\ref{einstein2}) for the scale factor $a$ it reduces to:
\begin{equation}
\dot H+3\Gamma H^2+{||^{(1)}A||^2\over 8a^6}(2-\Gamma)=0,
\label{e12}\end{equation}
a first integral of which is readily obtained:
\begin{equation}
\left(a^3\right)^.=\sqrt{\beta^4+4\tilde a^3a^{3(2-\Gamma)}}
\label{e13}\end{equation}
where $\tilde a(x^k)$ is an integration ``constant'' and where we have
set $\beta^2\equiv\left(3||{}^{(1)}A||^2/8\right)^{1/2}$. (We chose $\dot
a>0$ which will correspond to spacetimes emerging from a singularity.
The collapsing situation $\dot a<0$ is the time reversal of the
solution presented here. As for $\tilde a$ it will have to be positive
or zero:
see Eq (\ref{e23}) below.)
Eq (\ref{e13}) can be explicitely integrated when the anisotropy matrix
vanishes ($\beta^2=0$) or when matter
is  dust ($\Gamma=1)$, a
radiation fluid ($\Gamma=4/3$), or a stiff fluid ($\Gamma=2$)(the
particular case $\Gamma=0$ is treated at the end of the section); we
give the expression for $a$ in the case $\Gamma=4/3$ for completeness
only :
\begin{equation}
\Gamma=1 :\quad ^{(1)}a^3=u(\tilde a^3u+\beta^2)
\label{e14}\end{equation}
\begin{equation}
\Gamma=4/3:\quad u={3\beta^4\over16\tilde
a^{9/2}}\left(x\sqrt{x^2+1}-\ln{\left(\sqrt{x^2+1}+x\right)}\right)
\quad\hbox{with}\quad
x\equiv2\tilde a^{3\over2}({}^{(1)}a)/\beta^2
\end{equation}
\begin{equation}
\Gamma=2 :\quad ^{(1)}a^3=\bar a^3u\quad\hbox{with}\quad\bar
a^3\equiv\sqrt{\beta^4+4\tilde a^3}
\label{e16}\end{equation}
where $u\equiv t-t_0(x^k)$ with $t_0(x^k)$  an integration
``constant''. For a general
$0<\Gamma<2$ (and $\beta^2\neq 0$) an approximate solution for small $u$ is:
\begin{equation}
0<\Gamma<2\quad ^{(1)}a^3=\beta^2u\left[1+\tilde cu^{2-\Gamma}+{\cal
O}\left(u^{4-2\Gamma}\right)\right]\quad\hbox{with}\quad \tilde
c\equiv{2\tilde a^3\over3-\Gamma}\beta^{-2\Gamma}.
\label{e17}\end{equation}

\subsection{The metric}

The anisotropy matrix and the scale factor being known, Eq (\ref{e5}) then
gives the metric $^{(1)}\gamma_{ij}(t,x^k)$.  Indeed, assuming that
$^{(1)}A^i_j$ is diagonalizable (this does not spoil the genericity of
the solution), let us denote by $r_{(a)}(x^k)$ its three eigenvalues and
by $e_a$ three associated independent eigenvectors with components
$e^i_a(x^k)$ that we shall normalize to unity.  The triad $e_i^a$ forms
a basis of the tangent space.  Defining the cotriad $e_i^a$ by
$e^a_ie^i_b=\delta^a_b$, the components of any tensor on this new basis
is obtained by contracting its components in the coordinate basis with
the triad or cotriad.  In particular we define:
$\eta_{ab}\equiv\gamma_{ij}e^i_ae^j_b$ and $
\kappa_{ab}\equiv\kappa_{ij}e^i_ae^j_b$.  Since the triad is time
independent the relation $\kappa_{ab}=\dot\eta_{ab}$ holds and Eq
(\ref{e5}) becomes:  $\dot\gamma_{ab}=r_{(a)}\gamma_{ab}/a^3$ with
$\gamma_{ab}\equiv\eta_{ab}/a^2$.  Moreover the fact that
$A^k_i\gamma_{jk}=A^k_j\gamma_{ik}$ implies that $\eta_{ab}$ is diagonal
(see \cite{cdlp} for a demonstration).  Hence the solution for the
metric:
\begin{equation}
^{(1)}\gamma_{ij}=e_i^ae_j^b{}^{(1)}\eta_{ab}\quad\hbox{with}
\quad {}^{(1)}\eta_{ab}=\delta_{ab}
a^2\exp \left(r_{(a)}\int dt a^{-3}\right)
\label{e18}\end{equation}
where we recall that $r_{(a)}$ are the eigenvalues of the anisotropy
matrix  and $e^i_a$ three associated normalized eigenvectors (the
tracelessness of $A^i_j$ implies that $\Sigma r_{(a)}=0$; we also have
$\Sigma r_{(a)}^2=||{}^{(1)}A||^2=8\beta^4/3$). Explicit integration of Eq
(\ref{e18}) with $a$ given by Eq (\ref{e14}, \ref{e16}-\ref{e17}) gives:
\begin{equation}
\Gamma=1\quad
^{(1)}\eta_{ab}=\delta_{ab}C_{(a)}\beta^{2r_{(a)}/\beta^2}
u^{2/3+r_{(a)}/\beta^2}(u\tilde
a^3+\beta^2)^{2/3-r_{(a)}/\beta^2}
\label{e19}\end{equation}
\begin{equation}
\Gamma=2\quad
^{(1)}\eta_{ab}=\delta_{ab}\bar a^2C_{(a)}u^{2/3+r_{(a)}/\bar a^3}
\label{e20}\end{equation}
\begin{equation}
0<\Gamma<2\quad
^{(1)}\eta_{ab}=\delta_{ab}\beta^{4/3}C_{(a)} u^{2/3+r_{(a)}/\beta^2}
\left[1+\tilde
c\left({2\over3}-{r_{(a)}\over\beta^2(2-\Gamma)}\right)u^{2-\Gamma}+ {\cal
O}\left(u^{4-2\Gamma}\right)\right]
\label{e21}\end{equation}
where $C_{(a)}(x^k)$ are three integration ``constants''.
We have that ${\rm det}\gamma_{ij}=a^6={\rm det}^2(e^a_i){\rm det}\eta_{ab}
={\rm det}^2(e^a_i)C_{(1)}C_{(2)}C_{(3)}a^6$, and therefore
${\rm det}^2(e^a_i)C_{(1)}C_{(2)}C_{(3)}=1$.
We note that when expanding the metric (\ref{e19}) in small $u$, one finds
the metric (\ref{e21}) for $\Gamma=1$.
We also note that at leading order the metric (\ref{e21}), and therefore,
because of the previous remark, also the metric (\ref{e19}),
 is Kasner like. Indeed, setting $p_{(a)}=1/3+r_{(a)}/2\beta^2$,
it reads ${}^{(1)}\eta_{ab}\propto u^{2p_{(a)}}$ with $\Sigma
p_{(a)}=\Sigma p_{(a)}^2=1$. However it is important to note
that the metric (\ref{e20}) for $\Gamma=2$ is NOT Kasner-like, since the
sum of the $p_{(a)}^2$ (with the definition  $p_{(a)}=1/3+r_{(a)}/2\bar a^3$)
is less than 1 but one has still $\Sigma
p_{(a)}=1$.

Finally, we should mention that the previous calculations are only
valid for $\beta^2 \neq 0$. If $\beta^2=0$ then the anisotropy
matrix vanishes, and the integration of eq (\ref{e5}) is obvious.
 The metric is quasi-isotropic and reads
\begin{equation}
{}^{(1)}\gamma_{ij}=a^2(t) h_{ij}(x^k),
\end{equation}
where $h_{ij}$ is an arbitrary ``seed'' metric that depends only on space
and the scale factor must be taken from the paragraph b. depending on
which type of matter one considers. This particular case of a
quasi-isotropic metric was studied in details in our previous paper
\cite{cdlp}.

\subsection{The energy density}

 As for the energy density it is given by Eq (\ref{einstein3})
which reduces to:
\begin{equation}\chi\epsilon=3H^2-||{}^{(1)}A||^2/8a^6,
\label{e22}\end{equation}
that is, using (\ref{e13}) :
\begin{equation}
\chi\epsilon={4\over3}\tilde a^3a^{-3\Gamma}.
\label{e23}\end{equation}
(The positivity of $\epsilon$ implies that $\tilde a$ has to be
positive.) For ${}^{(1)}a$ given by Eq (\ref{e14}, \ref{e16}-\ref{e17}),
(\ref{e23}) yields:
\begin{equation}
\Gamma=1\quad \chi {}^{(1)}\epsilon={4\over3u(u+\beta^2/\tilde a^3)}
\label{e24}
\end{equation}
\begin{equation}\Gamma=2\quad\chi {}^{(1)}\epsilon={4\tilde
a^3\over4\tilde a^3+\beta^4}{1\over3u^2}\label{e25}\end{equation}
\begin{equation} 0<\Gamma<2\quad\chi
{}^{(1)}\epsilon={4\tilde
a^3\over3}\beta^{-2\Gamma}u^{-\Gamma}\left[1-\tilde c\Gamma u^{2-\Gamma}+{\cal
O}\left(u^{4-2\Gamma}\right)\right],
\label{e26}\end{equation}
and one notices that
the surface $u=0\Leftrightarrow t=t_0(x^k)$ is a singular surface of
infinite density. (Note also that if, in the general case $0<\Gamma<2$,
eq (\ref{e23}) yields (\ref{e26}), eq (\ref{e22}) only gives the
leading part in $\chi\epsilon$.)

\subsection{The three velocity}

As for the three velocity at first order it follows from (\ref{einstein4}).
However Eq (\ref{e10}), which at first order in the gradients reads
\begin{equation}
\tilde a^3a^{(3-3\Gamma)}\Gamma u_i=(1-\Gamma)\int\!
dt\, a^3\partial_i(\tilde a^3a^{-3\Gamma})+C_i,
\end{equation}
gives us the behaviour of $u_i$ without having to resort to the full
expression for the metric. In the case of dust it tells us for example
that $u_i$ is a function of space alone, and for $0<\Gamma<2$ it
gives:
\begin{equation}
{}^{(1)}u_i=\partial_it_0+\tilde C_iu^{\Gamma-1}[1+(\Gamma-1)\tilde c
u^{2-\Gamma}]+{1-\Gamma\over\Gamma}{\partial_i\tilde
c\over\tilde c}{u\over2-\Gamma},
\label{e28}\end{equation}
with $\tilde C_i=C_i\tilde a^{-3}\beta^{2(\Gamma-1)}/\Gamma$.
To determine the three ``constants'' $\tilde C_i(x^j)$ as functions of the
constants $C_{(a)}$ appearing in the metric, the more complete
eq (\ref{einstein4}) must be used;  their explicit expressions in the
case of spherical symmetry will be given in section 7.

\subsection{Genericity}

  Let us now examine the genericity of the metric thus obtained. It
depends on the following 12 arbitrary functions: $^{(1)}A^i_k(x^k)$ (8
functions) (or, equivalently: $r_{(a)}(x^k)$ (2 functions) and
$e^i_a(x^k)$ (6 functions));  $\tilde a(x^k)$ and $t_0(x^k)$;  and the
two functions $C_{(a)}(x^k)$.

Now 4 of these 12 functions can in principle be fixed by choosing a
particular synchronous reference frame (see section 7 for an explicit
implementation of such a gauge fixing in the case of spherical
symmetry).  One sees in particular that the reference frame can be
chosen in such a way that the surface of infinite density be $t=0$, that
is one can choose $t_0$ to be zero.  Indeed in an infinitesimal change
of coordinates:  $\tilde t=t+T$ and $\tilde x^i=x^i+X^i$ with $T=T(x^i)$
to preserve synchroniticity, the three velocity transforms as $\tilde
u_i={\partial t\over\partial x^i}u_0+{\partial x^j\over\partial
x^i}u_j\simeq -\partial_iTu_0+(\delta^j_i-\partial_iX^j)u_j\simeq
u_i-\partial_iT$ if $|u_i|<<1$.  Therefore the 3-velocity can be set
equal to zero by an appropriate choice of coordinates if it depends on
space only, which is the case for dust.  In all other cases this freedom
of gauge can be used to set $\partial_it_0=0$ as can be seen from Eq
(\ref{e28}).

The metric ${}^{(1)}\gamma_{ij}$ therefore depends on $12-4=8$
physically distinct arbitrary functions of space corresponding to the
4 degrees of freedom of gravity (the two gravitons) and the four
degrees of freedom of a fluid ($\epsilon$ and $u^i$). It is  therefore
generic.

\subsection{Late time limit}

The scale factor being an increasing function of time, Eq (\ref{e13}) tells
us that when $a$ is large the anisotropy $\beta^2$ becomes negligible
(unless $\Gamma=2$) and the scale factor tends to its
Friedmann-Robertson-Walker value
$a\propto u^{2/3\Gamma}\simeq t^{2/3\Gamma}$ ($t_0$ can be neglected
for large $t$). Since then $\int dt a^{-3}\propto
t^{(\Gamma-2)/\Gamma}\to 0$, Eq (\ref{e18}) tells us that
${}^{(1)}\gamma_{ij}$ tends
to a ``quasi-isotropic'' metric : ${}^{(1)}\gamma_{ij}\to
t^{4/3\Gamma}h_{ij}(x^k)$ where $h_{ij}(x^k)$ is a ``seed'' metric
depending on three physically distinct arbitrary functions. Five
physical degrees of freedom are therefore diluted away: the traceless
part of the intrinsic curvature and the epoch of the Big-Bang (in the
particular case $\Gamma=2$, the metric does not become quasi isotropic
at late times and only the epoch of the Big-Bang is lost). The
quasi-isotropic scheme developped within an Hamilton-Jacobi framework
by Salopek, Stewart and collaborators \cite{ss} and  along the lines
presented here in \cite{cdlp}, which consists in iterating Einstein's
equations starting from the restricted ``seed''
${}^{(1)}\gamma_{ij}=t^{4/3\Gamma}h_{ij}(x^k)$, is therefore justified
far away from a spacetime singularity. On the other hand, near a
singularity, the full first order metric must be taken as starting
point.

\subsection{The case of vacuum}

The Einstein equations for vacuum can be derived from the general
equations for a perfect fluid by imposing in (\ref{einstein1}-
\ref{einstein4}) $\epsilon=0$ and $S^i_j=0$.  Note that in this case all
the terms in (\ref{einstein2}) proportional to $\Gamma$ cancel because of
(\ref{einstein3}) with $\epsilon=0$.

In the vacuum case the constraint (\ref{einstein3}), that is (\ref{e23}),
gives $\tilde a=0$ and Eq (\ref{e13}) gives ${}^{(1)}a^3=\beta^2u$ so
that the metric (\ref{e18}) reads:
$\eta_{ab}=\delta_{ab}\beta^{4/3}C_{(a)}u^{2p_{(a)}}$ with
$p_{(a)}\equiv {\scriptstyle{1\over 3}} +r_{(a)}/2\beta^2$.  It is a
Kasner-like metric since
$\Sigma p_{(a)}=\Sigma p_{(a)}^2=1$.  (This incidently shows that matter
becomes negligible near $u=0$:  see Eq (\ref{e21}).) The metric depends
on 11 functions but (\ref{einstein4}) gives three additional
constraints.  Therefore the solution depends on 8 functions that is 4
physical degrees of freedom, those of the gravitons, as it should in
vacuum.

\subsection{The case of a cosmological constant}

In the particular case of a cosmological constant ($\Gamma=0$),
the integration of (\ref{e13}) gives:
\begin{equation}\beta^2\neq0:\quad
{}^{(1)}a^3={\beta^2\over\sqrt{3\Lambda}}\sinh(\sqrt{3\Lambda}\,u)\quad
,\quad\beta^2=0:\quad {}^{(1)}a^3=\exp(\sqrt{3\Lambda}\,u)
\end{equation}
with $\Lambda\equiv4\tilde a^3/3$ and the metric (\ref{e18}) reads:
$$\beta^2\neq0:\quad{}^{(1)}\eta_{ab}=\delta_{ab}\left(4\beta^4\over3
\Lambda\right)^{1/3}C_{(a)}(\sinh\sqrt{3\Lambda/4}\,
u)^{2/3+r_{(a)}/\beta^2}(\cosh\sqrt{3\Lambda/4}\,u)^{2/3-r_{(a)}/\beta^2}$$
\begin{equation}
\beta^2=0:\quad {}^{(1)}\gamma_{ij}={\rm
e}^{{\sqrt\Lambda\over3}\,t}h_{ij}(x^k).\end{equation}
(In the case $\beta^2=0$ the integration constant $t_0$ can be
absorbed in the seed metric $h_{ij}$.)
Eq (\ref{einstein3}) becomes a definition of the energy density:
\begin{equation}
\chi\,{}^{(1)}\epsilon=\Lambda=``Const.''
\end{equation}
and (\ref{e10}) says that $\Lambda$ is a true constant, independent of space,
and hence is not a true degree of freedom. The metric therefore
depends on 11 functions, 3 of which disappear when the constraint
(\ref{einstein4}) is imposed.  In that case then, as in vacuum, the
solution depends on 4 physical degrees of freedom, as it should.

\section{Conditions of validity}

The purpose of this section is to establish in which situations the
approximation scheme developped in the two previous sections is valid
and which remedy to give in the cases where it is not.  To check the
validity of the approximation scheme, we simply compare the third order
terms arising from the first iteration which were ignored up to now,
with the first order terms that we have just calculated.  Far from a
spacetime singularity when the first order metric reduces to its
quasi-isotropic component this was already done in \cite{ss} and
\cite{cdlp} with the conclusion that the next orders tend to zero as
time increases if matter violates the strong energy condition, i.e.  if
the fluid is ``inflationary''.  On the other hand, near a singularity
where the anisotropy matrix cannot be ignored, the next order, as we
shall see below, blows up generically as one approaches the singularity,
whatever equation of state matter satisfies, and we shall recover the
BKL (Belinski- Khalatnikov-Lifshitz) oscillatory behaviour for the
metric.

Let us begin with the most tiresome part:  the computation of the Ricci
tensor built from the first order three-dimensional metric.  We shall
here compute the Ricci tensor for a general metric of the form
\begin{equation}
\gamma_{ij}=e_i^ae_j^b\eta_{ab},
\end{equation}
where the triad $e^a_i$ depends only on the spatial coordinates whereas
the metric $\eta_{ab}$ depends both on spatial coordinates and time. The
situation is therefore more complicated than if $\eta_{ab}$ were only
time dependent as is the case in the BKL analysis.  However we proceed
along similar lines.

Let us first introduce the Ricci rotation coefficients (see \cite{schucking}),
 defined by
\begin{equation}
\gamma_{abc}\equiv e_{(a)i;k}e^i_{(b)}e^k_{(c)}, \end{equation}
and their commutator,
\begin{equation}
\lambda_{abc}\equiv \gamma_{abc}-\gamma_{acb},
\end{equation}
which have the
property
\begin{equation}
\gamma_{abc}=-\gamma_{bac}+\partial_c\eta_{ab},
\end{equation}
thus enabling us to express the Ricci rotation coefficients in terms of their
commutators:
\begin{equation}
\gamma_{abc}={\scriptstyle{1\over 2}}\left[\lambda_{abc}+\lambda_{bca}
-\lambda_{cab}
+\partial_c\eta_{ab}
+\partial_b\eta_{ac}-\partial_a\eta_{bc}\right].
\end{equation}
It is important to rewrite the commutators $\lambda_{abc}$ in the form
\begin{equation}
\lambda_{abc}=\eta_{ad}\mu^d_{\ bc}+\partial_c\eta_{ab}-\partial_b\eta_{ac},
\end{equation}
where
\begin{equation}
\mu^d_{\ bc}\equiv e^i_be^k_c\left(\partial_ke_i^d-\partial_ie_k^d\right),
\end{equation}
are {\it time independent}. In order to make the link with the
terminology of BKL it is worth noticing that these coefficients can be
rewritten in the form
\begin{equation}
\mu^a_{\ bc}=-{1\over \det(e)}\epsilon_{dbc}{\vec e}^d.
{\vec\nabla}\times{\vec e}^a.
\end{equation}

It is then straightforward to compute the components of the three dimensional
Riemann tensor in the nonholonomic basis. One finds
\begin{equation}
R_{abcd}=\eta_{ec}\eta_{fb}\partial_a\gamma_d^{\ ef}-\eta_{ec}\eta_{fa}
\partial_b\gamma_d^{\ ef} +\gamma_{deb}\gamma^e_{\ ca} -\gamma_{dea}
\gamma^e_{\ cb}
+\gamma_{dce}\gamma^e_{\ ba}-\gamma_{dce}\gamma^e_{\ ab}.
\end{equation}
Therefore the components  of the Ricci tensor are
\begin{equation}
R_{ab}=\eta_{eb}\partial_a\gamma_c^{\ ec}-\eta^{cd}\eta_{eb}\eta_{fa}
\partial_c
\gamma_d^{\ ef}+\gamma^c_{\ ec}\gamma^e_{\ ba}-\gamma^c_{\ be}\gamma^e_{\ ac} .
\end{equation}
In terms of the functions $\mu_{abc}$, the Ricci rotation coefficients
can be expressed as
\begin{equation}
\gamma_{abc}={\scriptstyle{1\over 2}}\left[\mu_{abc}+\mu_{bca}-\mu_{cab}
+\partial_c\eta_{ab}
+\partial_a\eta_{bc}-\partial_b\eta_{ac}\right].
\end{equation}
so that the explicit expression of the Ricci tensor in terms of the
triad and of the metric $\eta_{ab}$ is given by
\begin{eqnarray}
\lefteqn{R_{ab}=\partial_a\mu^c_{\ bc}-{\scriptstyle{1\over 2}}\partial_c
\mu^c_{\ ba}
-{\scriptstyle{1\over 2}}\eta_{eb}\eta^{cd}\partial_c\mu^e_{\ ad}
+{\scriptstyle{1\over 2}}\eta_{ea}\eta^{cd}
\partial_c\mu^e_{\ db}} \nonumber\\
&&+{\scriptstyle{1\over 2}}\mu^c_{\ ec}\left(\mu^e_{\ ba}+\mu_{ba}^{\ \ e}
-\mu^{\ e}_{a\ b}\right)
-{\scriptstyle{1\over 4}}\left(\mu^c_{\ ae}+\mu_{ae}^{\ \ c}-\mu_{e\ a}^{\ c}
\right)\left(\mu^e_{\ bc}+\mu_{bc}^{\ \ e}-\mu_{c\ b}^{\ e}\right) \nonumber\\
&&+{\scriptstyle{1\over 2}}\mu^c_{\ ec}\eta^{ed}\left(\partial_d\eta_{ab}
-\partial_a\eta_{bd}-\partial_b
\eta_{ad}\right)-{\scriptstyle{1\over 4}}\mu^e_{\ ba}
\eta^{cf}\partial_e\eta_{cf}
+{\scriptstyle{1\over 4}}\left(\mu_{ba}^{\ \ e}-\mu_{a\ b}^{\ e}\right)
\eta^{cf}
\left(2\partial_f\eta_{ec}-\partial_e\eta_{fc}\right) \nonumber\\
&&+{\scriptstyle{1\over 2}}\mu_{\ \ a}^{cd}\left( 2\partial_c\eta_{bd}
+\partial_d\eta_{bc}
-\partial_b\eta_{cd}\right)+{\scriptstyle{1\over 2}}\mu_a^{\ cd}
\partial_d\eta_{bc}
+{\scriptstyle{1\over 2}}\mu^{fe}_{\ \ b}\left(\partial_e\eta_{af}-
\partial_a\eta_{ef}\right)
-{\scriptstyle{1\over 2}}\mu_b^{\ fe}\partial_f\eta_{ea} \nonumber\\
&&+{\scriptstyle{1\over 4}}\eta^{ed}\eta^{cf}\left(\partial_d\eta_{ab}
-\partial_b\eta_{ad}
-\partial_a\eta_{bd}\right)\left(2\partial_f\eta_{ec}-\partial_e\eta_{cf}
\right)
+{\scriptstyle{1\over 4}}\eta^{ed}\eta^{cf}\partial_a\eta_{ef}\partial_b
\eta_{cd} \nonumber\\
&&+{\scriptstyle{1\over 2}}\eta^{cd}\eta^{ef}\left(\partial_c\eta_{ae}
\partial_d\eta_{bf}
-\partial_c\eta_{be}\partial_f\eta_{da}\right)
+{\scriptstyle{1\over 2}}\eta^{cd}\left(\partial_a\partial_d\eta_{bc}
-\partial_a\partial_b\eta_{dc}-\partial_c\partial_d\eta_{ab}
+\partial_b\partial_c\eta_{ad}\right).
\end{eqnarray}

In addition to the curvature terms we have also ignored, in the first
order approximation, the terms quadratic in the three-velocity.  The
approximate three-velocity that follows from the first order metric
according to (\ref{einstein4}) is given in the new basis by
\begin{equation}
 \chi\epsilon\Gamma u_a\simeq -{1\over 2}
\left(\partial_b\kappa^b_a+\kappa_d^b\gamma^d_{\ ab}-\kappa_a^d
\gamma^b_{\ db} -\partial_a \kappa\right),
\end{equation}
where we have used the relation
\begin{equation}
\kappa^j_{i;k}e^i_ae^b_je^k_c=\partial_c \kappa_a^b
+\kappa_d^b\gamma^d_{\ ac}-\kappa_a^d\gamma^b_{\ dc}.
\end{equation}

We have now to analyze the time dependence of these two expressions
giving the Ricci tensor and the three-velocity and find the dominant
terms, i.e.  that with the smallest power in $u$ since we are heading
towards the singularity.  To do that we shall assume that 1.  the new
metric is diagonal, i.e.  $\eta_{ab}=\eta_a\delta_{ab}$ (this amounts to
supposing that the anisotropy matrix is diagonalizable) and 2.  the
spatial derivative of any component of the metric has the same time
behaviour than the component itself (this means in particular that we
assume that $t_0$ is independent of space -- which, as we saw, can be
the case in an appropriate reference frame --, and that we also ignore
the logarithmic corrections that arise from the spatial derivative of
the exponents).  Inspection of the above expression then shows that all
the terms in $R_{ab}$ can be classified in one of the following
categories, as far as their time behaviour is concerned:  cste (or
logarithmic), $\eta_a/\eta_c$, $\eta_b/\eta_c$, $\eta_c/\eta_d$ and
$\eta_a\eta_b/\eta_c\eta_d$, where $a$ and $b$ are fixed but $c$ and $d$
range from 1 to 3, whereas the three-velocity, because simpler can be
given explicitly here:
\begin{equation}
\chi\epsilon\Gamma {}^{(1)}u_a\simeq -{\scriptstyle{1\over 2}}
\left[\Sigma_{b\neq a} {\dot\eta_b\over \eta_b}
\gamma^b_{\ ab}-{\dot\eta_a\over \eta_a}\gamma^b_{\ ab}
-\Sigma_{b\neq a}\partial_a\left({\dot\eta_b\over \eta_b}\right)\right],
\end{equation}
with
\begin{equation}
\gamma^b_{\ ab}=\mu^b_{\ ab}+{\scriptstyle{1\over 2}}
\left(\partial_b\eta_{ac}+\partial_c\eta_{ab}
-\partial_a\eta_{bc}\right).
\end{equation}

\subsection{General case}

Now, as was shown in section 3, the generic behaviour of the first order
metric near the singularity is of Kasner type, when $0<\Gamma<2$.  Let
us label the coordinates in such a way that $p_1<p_2<p_3$.  We know that
$-1/3<p_1<0$, $0<p_2<2/3$ and $2/3<p_3<1$ (see \cite{ll}).  In $\tilde
R_{11}$, the dominant term then is
\begin{equation}
{\scriptstyle{1\over 2}}\left(\mu^1_{\ 23}\right)^2\eta_1^2/\eta_2\eta_3
\end{equation}
(there is no term of the form $\eta_1^2/\eta_3^2$ because of the
antisymmetry of $\mu^a_{\ bc}$ in the two last indices).
In $\tilde R_{22}$, the dominant term is
\begin{equation}
-{\scriptstyle{1\over 2}}\left(\mu^1_{\ 23}\right)^2 \eta_1/\eta_3
\end{equation}
whereas the dominant term in $\tilde R_{33}$ is
\begin{equation}
-{\scriptstyle{1\over 2}}\left(\mu^1_{\ 23}\right)^2 \eta_1/\eta_2.
\end{equation}
As for the dominant contribution in the crossed terms, $\tilde R_{12}$,
$\tilde R_{23}$ and $\tilde R_{31}$, it is more complicated since
they are several terms involved. Therefore we quote only the time
dependence,
\begin{equation}
\tilde R_{12}\sim \eta_1/\eta_3, \qquad \tilde R_{23}\sim cst, \qquad
\tilde R_{31}\sim\eta_1/\eta_2.
\end{equation}

The dominant term in the scalar three curvature $\tilde R$
is
\begin{equation}
-{\scriptstyle{1\over 2}}\left(\mu^1_{\ 23}\right)^2 {\eta_1\over\eta_2\eta_3}.
\end{equation}
The three curvature thus
 behaves as a power law,
\begin{equation}
\tilde R\sim u^{2(p_1-p_2-p_3)}\sim u^{4p_1}u^{-2}.
\end{equation}
As we see the dominant terms always come from the cross-product of the
$\mu_{abc}$ so that the terms with the spatial derivatives of the metric
$\eta_{ab}$ do not play a role near the singularity.  One can therefore
expect that we will recover the results obtained by BKL who started
their analysis on a Bianchi IX model where the metric $\eta_{ab}$ is
only time dependent.

The time behaviour of the Ricci tensor being now known, let us see if we
were allowed to neglect it.  An analysis of the Einstein equations
(\ref{einstein1}), (\ref{einstein2}) and (\ref{einstein3}), shows first
that $u^2\tilde R_i^j$ and $\tilde R u^2$ must be convergent for the
approximation to be valid.  This is NOT the case if the dominant terms
are those listed above because $p_1$ is negative.  Indeed $u^2 \tilde R$
is divergent as well as $u^2\tilde R_1^3$ and $u^2\tilde R_1^2$.

As for the term containing the three-velocity in the equation
(\ref{einstein2}), it goes like
\begin{equation}
u^2(\epsilon \Gamma u^au_a)\sim u^{\Gamma-2p_3}.
\end{equation}
Therefore this term is convergent only if $\Gamma>2p_3$. Moreover, since
$-1/3<p_1<0$ and $2/3<p_3<1$, one can see that when this term diverges
 it can be either more  ($\Gamma <4p_1+2p_3$) or less divergent than
the curvature term.

The conclusion therefore is that the long wavelength approximation scheme
 breaks down in the general case when approaching the singularity.

\subsection{Case $\mu^1_{\ 23}=0$ }

In the particular case where
\begin{equation}
\mu ^1_{\ 23}=0,
\label{cond}\end{equation}
all the dominant contributions listed above vanish and the validity of
the scheme must be reconsidered.  In that case the dominant time
dependences are the following:
\begin{equation}
\tilde R_{11}\sim \eta_1/\eta_3, \tilde R_{22}\sim \eta_2/\eta_3,
\tilde R_{33}\sim cst,
\end{equation}
\begin{equation}
\tilde R_{12}\sim \eta_2/\eta_3, \tilde R_{23}\sim cst,
\tilde R_{31}\sim cst.
\end{equation}
Therefore
\begin{equation}
\tilde R\sim 1/\eta_3\sim u^{-2p_3}.
\end{equation}
Knowing that $2/3<p_3<1$, one can conclude that $\tilde R u^2$ is then
always convergent.  In a similar manner one sees that all the quantities
$u^2\tilde R_i^j$ are convergent.

Let us now look at the term quadratic in the velocity.  The time
dependence of this term remains the same as in the general case.
Therefore the approximation of equation (\ref{einstein2}) is valid only
(except the case $\Gamma=4/3$) if $\Gamma>2p_3$ or in the case where
$u_3=0$ if
$\Gamma>2p_2$ or $u_2=0$, these conditions being rather restrictive.

This is however not yet the end of the story.
The analysis of eq (\ref{einstein3}) imposes two further conditions:
\begin{enumerate}
\item $u^au_a$ must be small with respect to 1, which implies that
$\Gamma>1+p_a$ for all $a$ unless $u_a$ vanishes.
\item   $\tilde R$ must be negligible with
respect to the energy density $\epsilon$:  this is due to the fact,
already mentionned, that the two first terms on the right hand side
compensate at leading order.  The condition is therefore, in view of
equations (\ref{e24}-\ref{e26}), that $u^\Gamma \tilde R$ must be
convergent.  Therefore one must have $\Gamma>2p_3$.
\end{enumerate}
Note that the condition $\Gamma>1+p_3$ implies all the other conditions,
but this is very restrictive in general since $p_3$ is limited from
above only by $1$.

To summarize, we find that the first order solution given in section 3
is a good approximation to a solution of Einstein's equations near a
spacetime singularity, if the conditions $\mu^1_{\ 23}=0$, $\Gamma>2p_3$ and
$\Gamma>1+p_a$ (for all $a=1,2,3$ unless $u_a=0$), are satisfied.  Diagonal
anisotropy matrices form an important subclass of matrices satisfying
the first condition and we shall see that in the context of spherical
symmetry, studied in detail in the last section, the other conditions
are also fulfilled for $\Gamma=1$ or for $\Gamma>4/3$.  Now imposing
(\ref{cond}) renders the first order solution non generic as it then
depends on 7 instead of 8 physically distinct arbitrary functions.
However a qualitative analysis of what happens when $\mu^1_{\ 23}\neq0$
can be given, which follows closely the work of BKL.

\subsection{BKL oscillatory behaviour}

We now consider the Einstein equations (\ref{einstein1}-\ref{einstein2})
where we do NOT neglect the curvature terms any longer.  The time
derivative of equation (\ref{extr}), after use of (\ref{einstein2}) and
(\ref{einstein3}) and expressed in the new basis, gives
\begin{equation}
a^{-3}\left(a^3\kappa^a_b\right)\dot{}=(2-\Gamma)\chi\epsilon\delta^a_b
+2\chi\epsilon \Gamma u^au_b-2\tilde R^a_b. \label{bkl}
\end{equation}
As shown in section 3, the energy density $\epsilon$ goes like
$u^{-\Gamma}$. We now assume that
the curvature term evolving like $u^{4p_1-2}$ is dominant over the velocity
term, evolving like $u^{\Gamma-2-2p_3}$ (at worse).  Assume moreover that
the metric remains diagonal in the evolution and that at some time it is
Kasner-like.  In these conditions we are able to recover the behaviour
discovered by BKL, initially in the case of Bianchi IX and extended
later to inhomogenous situations.

Following BKL, let us introduce a new time defined by
\begin{equation}
\tau=\ln u,
\end{equation}
and let us write the metric in the form
\begin{equation}
[\eta_{ab}]=Diag[e^{2\alpha},e^{2\beta},e^{2\gamma}].
\end{equation}
Then keeping in (\ref{bkl}) only the dominant contributions from the
curvature term, we get the three following equations governing the
coefficients of the diagonal metric,
\begin{equation}
\partial_\tau^2\alpha=-{\scriptstyle{1\over 2}} \left(\mu^1_{\ 23}\right)^2
e^{4\alpha}, \quad
\partial_\tau^2\beta={\scriptstyle{1\over 2}}
\left(\mu^1_{\ 23}\right)^2e^{4\alpha},
\quad
\partial_\tau^2\gamma={\scriptstyle{1\over 2}}\left(\mu^1_{\ 23}\right)^2
e^{4\alpha}.
\end{equation}
When the metric is Kasner-like, one has
\begin{equation}
\partial_\tau\alpha=p_1, \qquad \partial_\tau\beta=p_2, \qquad
\partial_\tau\gamma=p_3,
\end{equation}
The equation for $\alpha$ is similar to the equation for a particle with
coordinate $\alpha$ moving in an exponential potential. Initially the
particle moves with a constant velocity $\partial_\tau\alpha=p_1$. After
reflexion on the potential wall, the particle will move with the velocity
$\partial_\tau\alpha=-p_1$. The two other equations then give the two other
final velocities $p_1$: $\partial_\tau\beta=p_2+2p_1$ and
$\partial_\tau\gamma=p_3+2p_1$.
Therefore the initial Kasner-like metric evolves into another
Kasner-like metric due to the influence of the curvature terms, given by
\begin{equation}
\eta_{ab}\sim Diag[u^{-p_1\over 1+2p_1},u^{p_2+2p_1\over 1+2p_1},
u^{p_2+2p_1\over 1+2p_1}]
\end{equation}
We thus recover from our general analysis the oscillatory behaviour
between Kasner-like metrics, behaviour which was studied in details
by BKL (see e.g. \cite{ll}).

Let us conclude this section with the particular cases of the vacuum,
a cosmological constant and a stiff fluid.  In the case of vacuum, one
has still a Kasner-like metric as was shown in the subsection 3.h.
Therefore the above analysis applies without modification.  The same
conclusion arises from the cosmological constant case with anisotropy
since the metric ( see subsection 3.i), when $u\rightarrow 0$ has the
same Kasner-like behaviour as the metric (\ref{e21}).  However the case
$\Gamma=2$ gives a qualitatively different result.
Indeed, there is more freedom for the
coefficients $p_{(a)}$ and it is then possible that $p_1$, the smallest
of the three powers, be positive, in which case all the terms $u^2
\tilde R_i^j$ converge.  Note that the energy equation is valid also
only in the case where $p_1>0$.  BLK showed that in the case of stiff
matter (or scalar field) the oscillating behaviour (if there exists one
at the beginning, i.e.  if $p_1$ is negative) will end after a few
oscillations when one goes backwards in time.  This means that,
sufficiently near the singularity, the approximation scheme works.

\section{The generic third order solution}

Let us first  rewrite the Einstein equations (\ref{einstein1}-
\ref{einstein4}) in the new basis:
\begin{equation}
\dot A^a_b=2a^3(\bar S^a_b-\bar R^a_b)
\end{equation}
\begin{equation}
2\dot H+3\Gamma H^2+{||A||^2\over8 a^6}(2-\Gamma)=
{\scriptstyle{1\over6}}(2-3\Gamma)\tilde R
-\scriptstyle{1\over 3}\chi\epsilon(4-3\Gamma)\Gamma u^au_a
\end{equation}
\begin{equation}
\chi\epsilon(1+\Gamma u^au_a)=3 H^2-{||A||^2\over8a^6}
+\scriptstyle{1\over 2}\tilde R
\label{neinstein3}
\end{equation}
where $||A||^2\equiv A^a_bA^b_a$, and
\begin{equation}
 \chi\epsilon\Gamma u_a=-{1\over 2\sqrt{1+u^cu_c}}
\left(\partial_b\kappa^b_a+\kappa_d^b\gamma^d_{\ ab}-\kappa_a^d
\gamma^b_{\ db} -\partial_a \kappa\right). \label{neinstein4}
\end{equation}

In the previous sections we have only considered the first order
approximation of the Einstein equations. We can now include third order
corrections to the first order quantities:
\begin{equation}
a={}^{(1)}a+{}^{(3)}a+\dots, \quad A^a_b={}^{(1)}A^a_b+{}^{(3)}A^a_b+\dots.
\end{equation}
These third order corrections follow from  the approximate Einstein equations
in which the terms that were ignored previously are now taken into
account but are computed with the first order solution. Since the third
order solution is supposed to be small with respect to the first order
solution, one can linearize the Einstein equations and all the equations
giving the third order terms will be linear ordinary differential equations.
One sees that, in principle, one can repeat this procedure at any order
and build iteratively the metric and the other quantities.

It is convenient to define the following third order quantities:
\begin{equation}
a_3\equiv {{}^{(3)}a\over {}^{(1)}a}
\end{equation}
and
\begin{equation}
{\bf y_3}\equiv {}^{(3)}\left({{\bf A}\over a^3}\right)={{}^{(3)}{\bf A}
\over ({}^{(1)}a)^3}
-3{}^{(1)}\left({{\bf A}\over a^3}\right)a_3,
\end{equation}
where a boldfaced letter stands for a matrix.
The third order correction to the extrinsic curvature, ${}^{(3)}\kappa_a^b$,
 can be decomposed, following (\ref{extr}), into
\begin{equation}
{}^{(3)} \kappa_a^b=2 {}^{(3)}H \delta_a^b+({\bf y_3})_a^b,
\end{equation}
where
\begin{equation}
{}^{(3)}H=\dot a_3.
\end{equation}
The expansion of Einstein equations then gives
\begin{equation}
{}^{(3)}A^a_b=2\int dt\  {}^{(1)}a^3(\bar S^a_b-\bar R^a_b) ,
\label{3einstein1}\end{equation}
where the term under the integral is built from the first order solution
and not from the exact solution as is the case in the exact Einstein
equation (\ref{einstein1}) (Note also that the constant of integration
that could arise from the above integral is supposed to be already included
in the first order term ${}^{(1)}A^a_b$).  As for the equations for $a$
and $\epsilon$ they become:
\begin{equation}
2\ddot a_3+6\Gamma H \dot a_3+{2-\Gamma\over 4}{}^{(1)}\left({{\bf A}
\over a^3}\right).{\bf y_3}={2-3\Gamma\over 6}\tilde R
-{\Gamma(4-3\Gamma)\over 3}\chi{}^{(1)}\epsilon u^a u_a.
\label{3einstein2}\end{equation}

The third order correction to the metric can be computed by expanding
formula (\ref{extr}) (in the new basis).  One finds
\begin{equation}
{}^{(3)}\kappa_a^b={}^{(3)}\dot\eta_{ac}\eta^{cb}-\dot\eta_{ac}\eta^{ce}
{}^{(3)}\eta_{de}\eta^{bd}.
\end{equation}
Using the fact that the matrix $[\eta_{ab}]$ is diagonal,
\begin{equation}
{}^{(1)}\eta_{ab}=Diag[\eta_c],
\end{equation}
one finds
\begin{equation}
{}^{(3)}\kappa_a^b={}^{(3)}\dot\eta_{ab}\eta_b^{-1}-\dot\eta_a\eta_a^{-1}
{}^{(3)}\eta_{ab}\eta_b^{-1},
\end{equation}
where there is no summation on the indices.
This equation can be integrated into
\begin{equation}
{}^{(3)}\eta_{ab}=\eta_a\int dt {\eta_b\over \eta_a}{}^{(3)}\kappa_a^b.
\end{equation}
and therefore
\begin{equation}
{}^{(3)}\eta_{ab}=\eta_a\int dt {\eta_b\over \eta_a}\left(2{}^{(3)}H
\delta^b_a +\left({\bf y_3}\right)_a^b\right).
\end{equation}
One can then obtain the third order energy and velocity by inserting the
metric ${}^{(1)}\eta_{ab}+{}^{(3)}\eta_{ab}$ in equations (\ref{neinstein3})
and (\ref{neinstein4}) (taking for the quadratic term $u_au^a$ the
first order solution ${}^{(1)}u_a$). For instance the third order
energy density is given by
\begin{equation}
\chi {}^{(3)}\epsilon=6 {}^{(1)}H\dot a_3-{1\over 4}{}^{(1)}
\left({{\bf A}\over a^3}\right).{\bf y_3}
+\scriptstyle{1\over 2}\tilde R-\chi \Gamma{}^{(1)}
\epsilon {}^{(1)}u^a{}^{(1)}u_a.
\end{equation}

\section{The example of spherical symmetry}
\subsection{The equations}

The line element of a spherically symmetric spacetime can be written
in a suitable coordinate system ($t,r,\theta,\phi$) as:
\begin{equation}
 ds^2=-dt^2+\gamma_{rr}dr^2+\gamma_{\theta\theta}d\Omega^2\quad\hbox{with}
\quad d\Omega^2\equiv d\theta^2+\sin^2\theta d\phi^2.
\label{s1}\end{equation}
An infinitesimal change of coordinates $\tilde t=t+T, \tilde r=r+R$ preserves
the synchronicity of the reference frame if $T=T(r)$ and $\dot
R=T'/\gamma_{rr}$ (where a prime denotes a derivative with respect to
$r$). It involves two arbitrary functions of space:
$T(r)$ and the integration ``constant'' in the equation for $R$  (see e.g.
\cite{ll}).

 The extrinsic curvature $\kappa^i_j\equiv\gamma^{ik}\dot\gamma_{jk}$
is diagonal and therefore the matrix $A^i_j$ in Eq (4) is too, so that we are
spared from the triad formalism of sections 3-5. We shall give here the
solution to third order in the  gradients. This will illustrate the general
discussion of the  preceeding sections, will clarify the gauge issue and allow
a
comparison with the known exact solution of Tolman-Bondi. This section is
intended to be self-contained.

When the line element is (\ref{s1}) so that the traceless anisotropy
matrix $A^i_j$ in $\kappa^i_k\equiv 2H\delta^i_j+A^i_j/a^3$ (where
$H\equiv\dot a/a$) is diagonal with eigenvalues $r_{(r)}=-A$,
$r_{(\theta)}=r_{(\phi)}=A/2$, the Einstein's equations
(\ref{einstein1}-\ref{einstein4}) for a perfect fluid yield
\begin{equation}
\gamma_{rr}=a^2\exp -\int\!dt\, {A\over a^3}\quad ,\quad
\gamma_{\theta\theta}=a^2\exp\left[ {\scriptstyle{1\over2}}\int\!dt\,
{A\over a^3}\right]
\label{s2}\end{equation}
where $A$ and $a$ are given by:
\begin{equation}
A=-2\int\!dt\,a^3(\bar S-\bar R)\quad\hbox{with}\quad\bar
S={2\chi\epsilon\over3}\Gamma u^ru_r\quad,\quad \bar R=\tilde
R^r_r-{\scriptstyle{1\over3}}\tilde R
\label{s3}\end{equation}
\begin{equation}
2\dot H+3\Gamma H^2+{3\over16}\left({A\over
a^3}\right)^2(2-\Gamma)={\scriptstyle{1\over6}}(2-3\Gamma)\tilde
R-{\scriptstyle{1\over2}} (4-3\Gamma)\bar S.
\label{s4}\end{equation}
As for the energy density and the radial velocity they are given by:
\begin{equation}
\chi\epsilon(1+\Gamma u^ru_r)=3H^2-{3\over16}\left({A\over a^3}\right)^2+
{\scriptstyle{1\over2}}\tilde R
\label{s5}\end{equation}
\begin{equation}
\chi\epsilon\Gamma u_r\sqrt{1+\Gamma u^ru_r}=2H'+{\scriptstyle{1\over2}}
\left({A\over a^3}\right)'+{\scriptstyle{3\over4}}\left({A\over a^3}\right)
(\ln\gamma_{\theta\theta})'.
\label{s6}\end{equation}
Simple counting gives that a generic metric solution of (\ref{s2}-\ref{s4})
depends on 4 arbitrary functions of $r$.  The total number of physical
degrees of freedom however is 2 (there are no gravitons in spherically
symmetric spacetimes and the fluid is specified by its density and
radial velocity).  Two functions can therefore be eliminated by fixing
the gauge that is choosing a particular synchronous reference frame, in
agreement with the remark below Eq (\ref{s1}).  We can first give a
geometrical meaning to the coordinate $r$ by relating it to the surface
of 2-spheres:  this will fix a function in $\gamma_{\theta\theta}$.  To
eliminate the remaining gauge freedom we note that in an infinitesimal
change of coordinates that preserves synchronicity, the radial velocity
transforms as:  $\tilde u_r={\partial t\over\partial\tilde
r}u_0+{\partial r\over\partial\tilde r}u_r\simeq
-T'u_0+\left(1-{\partial R\over\partial r}\right)u_r\simeq u_r -T'$ if
$u_r<<1$, so that an arbitrary function of space ($T'$) can be
substracted to $u_r$.

\subsection{The case of dust ($\Gamma=1$)}

At first order in the gradients Eq (\ref{s3}) gives $A=A(r)$ and
the solution of (\ref{s4}) is:
\begin{equation}
{}^{(1)}a^3=\tilde a^3u(u+\alpha)\qquad\hbox{with}\qquad u\equiv t-t_0
\quad\hbox{and}\quad\alpha\equiv {\scriptstyle{3\over4}}{|A|\over\tilde a^3}
\end{equation}
and where $\tilde a(r)$ and $t_0(r)$ are 2 integration ``constants".
The metric then follows from (\ref{s2}) ---see eq (\ref{e19}):
\begin{equation}
{}^{(1)}\gamma_{rr}=C_ru^{4/3}(1+\alpha/u)^{2/3(1+2\epsilon)}\qquad ; \qquad
{}^{(1)}\gamma_{\theta\theta}=C_\theta u^{4/3}(1+\alpha/u)^{2/3(1-\epsilon)}
\end{equation}
where $\epsilon\equiv A/|A|$. From now on we shall consider only
$\epsilon=1$ since the metric, in the case $\epsilon=-1$, tends toward
a metric of the type Kasner with the coefficients $(p_1=1, p_2=0, p_3=0)$
 which is nothing less than the flat metric as can be shown with a suitable
change of coordinates (see \cite{ll}): therefore there is no
singularity for $\epsilon=-1$.  $C_r$ and $C_\theta$ are 2 integration
constants.  The first order metric depends, as anticipated, on 4
functions of $r$:  $\alpha, t_0, C_r$ and $C_\theta$ and 2 can be
eliminated by fixing the gauge.  To do that we first impose
$C_\theta=r^2$, so that the radial velocity (\ref{s6}) becomes:
\begin{equation}
{}^{(1)}u_r=t'_0+{3\over2 r}\alpha
\label{s12}\end{equation}
which, as we already knew from (\ref{e28}) depends on space only and
can be set equal to zero (indeed when matter is dust and hence follows
geodesics there exists a synchronous reference frame where the particles
remain at rest) by choosing $\alpha=-2t_0'r/3$.

The gauge being thus completely fixed, the first order spherical
symmetric line element for dust finally reads:
\begin{equation}
{}^{(1)}ds^2=-dt^2+{\varrho'^2\over1-hr^2}dr^2+\varrho^2d\Omega^2
\label{s13}\end{equation}
with the definitions:  $\varrho\equiv ru^{2/3}$, $u\equiv t-t_0$ and
$C_r\equiv (1-hr^2)^{-1}$ (this form will be useful for a comparison
with the exact solution of Tolman and Bondi).  It depends on 2 arbitrary
functions of $r$:  $h$ and $t_0$.  The energy density follows from
(\ref{s5}) (see Eq (\ref{e24})) and the radial velocity is zero:
\begin{equation}
\chi\ {}^{(1)}\epsilon={4\over3u(u-2t'_0r/3)}\qquad ,\qquad {}^{(1)}u_r=0.
\end{equation}
Useful secondary quantities are:
\begin{equation}
{}^{(1)}a^3=\tilde a^3u(u-2t'_0r/3)\qquad ;\qquad {}^{(1)}\left({A\over
a^3}\right)={-8t'_0r\over9u(u-2t'_0r/3)}.
\label{s15}
\end{equation}

To obtain the third order metric the easiest way is as follows: Writing $a=
{}^{(1)}a(1+a_3); A/a^3={}^{(1)}(A/a^3)+y_3$, Eq (\ref{s2}) gives:
\begin{equation}
{}^{(1)}ds^2+{}^{(3)}ds^2=-dt^2+{\varrho'^2\over1-hr^2}(1+\gamma_r)dr^2+
\varrho^2(1+\gamma_\theta)d\Omega^2 \label{s16}
\end{equation}
with $\gamma_r=2a_3-\int\!dt\,y_3$, $\gamma_\theta=2a_3+{\scriptstyle{1\over2}}
\int\!dt\,y_3$. Eq (\ref{s4}) for $a_3$ and $y_3$ is then transformed,
by using the relation $\dot y_3=-3Hy_3-3{}^{(1)}(A/a^3)\dot a_3-
2(\bar S-\bar R)$ which follows from (\ref{s3}), into an equation for
$\gamma_\theta$ which eventually reads:
\begin{equation}
\ddot\gamma_\theta+3\dot\gamma_\theta \left[{}^{(1)}H+ {\scriptstyle{1\over4}}
{}^{(1)}({A/a^3})\right]-\bar R+ {\scriptstyle{1\over6}}\tilde R=0.
\label{s17}\end{equation}
With ${}^{(1)}H\equiv{}^{(1)}\dot a/a$ and ${}^{(1)}({A/a^3})$ given by
(\ref{s15}), and the relevant components of the Ricci tensor for the
line element (\ref{s13}) being:
\begin{equation}
\bar R={(hr^2)'\over3\varrho\varrho'}-{2hr^2\over3\varrho^2}\quad ;\quad
\tilde R={2hr^2\over\varrho^2}+{2(hr^2)'\over\varrho\varrho'},
\end{equation}
Eq (\ref{s17}) becomes $\ddot\gamma_\theta+2\dot\gamma_\theta /u+hu^{-4/3}=0$
which is readily integrated into:
\begin{equation}
\gamma_\theta=-{9\over10}hu^{2/3}.
\label{s19}\end{equation}
Turning then Eq (\ref{s4}) into an equation for $\gamma_r$ instead of
$\gamma_\theta$ and using the relation $\dot\gamma_r=\dot\gamma_\theta-
{\scriptstyle{3\over2}}y_3$ to eliminate $y_3$ one gets:
\begin{equation}
\ddot\gamma_r+3\dot\gamma_r\left[{}^{(1)}H- {\scriptstyle{1\over4}}
{}^{(1)}({A/a^3})\right]+2\bar R+ {\scriptstyle{1\over6}}\tilde R-
 {\scriptstyle{3\over4}}\dot\gamma_\theta {}^{(1)}({A/a^3})  =0
\label{s20}\end{equation}
the integration of which yields:
\begin{equation}
\gamma_r=-{9\over10}(h+h'r){u^{5/3}\over u-2t'_0r/3}+{6\over5}hrt'_0
{u^{2/3}\over u-2t'_0r/3}.
\label{s21}\end{equation}
At third order then, the spherically symmetric element for dust is (\ref{s16})
with
$\gamma_\theta$ and $\gamma_r$ given respectively by
(\ref{s19}) and (\ref{s21}).

Now the exact solution for dust with spherical symmetry is known. It is the
Tolman-Bondi solution, the line element of which can be written as (see
e.g. Landau Lifschitz \cite{ll}):
\begin{equation}
ds^2=-dt^2+{\rho'^2\over 1+f(r)}dr^2+\rho^2d\Omega,
\end{equation}
with
\begin{equation}
\rho={\mu\over 2f}\left(\cosh\eta-1\right), \qquad t-t_0(r)={\mu\over
 2f^{3/2}}\left(\sinh\eta-\eta\right)
\end{equation}
for $f>0$,
\begin{equation}
\rho={\mu\over -2f}\left(1-\cos\eta\right), \qquad t-t_0(r)={\mu\over
 2(-f)^{3/2}}\left(\eta-\sin\eta \right)
\end{equation}
for $f<0$, and
\begin{equation}
\rho={9\mu\over 4}^{1/3}\left(t-t_0(r)\right)^{2/3}
\end{equation}
for $f=0$.
The line element is written in the comoving gauge where the three velocity
vanishes and the miscalleneous functions appearing in the metric have an easy
physical interpretation: any particle is labelled by the coordinate $r$, the
same at any time; $4\pi\rho^2(r,t)$ gives the area of the sphere
containing this particle at time $t$; $\dot\rho(r,t)$ is the radial
velocity of the particle and $\mu (r)$ corresponds to the mass inside
the sphere containing the particle.

We now consider the expansion of the Tolman-Bondi solution in the  parameter
$u=t-t_0(r)$, supposed to be small. We find for $\rho$:
\begin{equation}
\rho=\left(3\over 2\right)^{2/3}\mu^{1/3}u^{2/3}+{9\over 20}
\left(2\over 3\right)^{2/3}\mu^{-1/3}fu^{4/3}.
\end{equation}
In the derivation of this expansion, we assume that $u'$ is of the order of
$u$. Going to the next to leading order in the expansion we thus find :
\begin{equation}
\gamma_{rr}={1\over 9}\left(3\over 2\right)^{4/3}{(\mu u^2)^{2/3}\over 1+f}
\left[\left(\ln\mu u^2\right)'\right]^2
\left[1+{3fu^{2/3}\over 5\mu^{2/3}}\left(2\over 3\right)^{1/3}
\left(\ln\mu u^2\right)'
{\left(\ln{f^3u^4/\mu}\right)'\over\left(\ln\mu u^2\right)'} \right].
\label{s27}\end{equation}
Identifying $f(r)=-hr^2$ and $\mu=4r^3/9$, we recover the first and
third orders (\ref{s16},\ref{s19}) and (\ref{s21}) given by the
expansion scheme.

\subsection{The general case  ($0<\Gamma<2$)}

The first order metric is obtained as before and, without going  into
details, it should be clear that it is given by (\ref{e21}), the index
$(a)$ being $r$ and $\theta$, with $r_{(r)}=-A=-4\epsilon\beta^2/3;
r_{(\theta)}=r_{(\phi)}=A/2$ (again we shall consider only $\epsilon=+1$
since $\epsilon=-1$ does not describe a singularity).  This metric
depends on 4 functions of $r$:  $C_r\beta^{4/3}=g(r)$,
$C_\theta\beta^{4/3}$, $t_0$ and $\tilde c$ and 2 of those can be
eliminated by particularizing the reference frame.  As before we shall
first choose $C_\theta\beta^{4/3}=r^2$.  Then, from (\ref{e28}) and the
discussion in section 3.f, we know that $t_0$ can be chosen to be zero,
and that fixes the gauge completely (note that when $\Gamma=1$ it is not
the comoving gauge chosen in the preceeding paragraph).  The first order
line element therefore is (\ref{s1}) with:
\begin{equation}
{}^{(1)}\gamma_{rr}=gt^{-2/3}\left[1+{2\tilde c\over3}
{(4-\Gamma)\over(2-\Gamma)}t^{2-\Gamma}+{\cal O}\left(t^{4-2\Gamma}\right)
\right]
\label{s28}\end{equation}
\begin{equation}
{}^{(1)}\gamma_{\theta\theta}=r^2t^{4/3}\left[1+{2\tilde c\over3}
{(1-\Gamma)\over(2-\Gamma)}t^{2-\Gamma}+{\cal O}\left(t^{4-2\Gamma}\right)
\right]
\label{s29}\end{equation}
which depends on 2 arbitrary functions: $g(r)$ and $\tilde c(r)$. As for the
density and radial velocity they are given by (\ref{e26}) and (\ref{s6}):
\begin{equation}
\chi {}^{(1)}\epsilon={\scriptstyle{2\over3}}(3-\Gamma)\tilde ct^{-\Gamma}
\left[1-\tilde c\Gamma t^{2-\Gamma}+{\cal O}\left(t^{4-2\Gamma}\right)\right]
\label{s30}\end{equation}
\begin{equation}
{}^{(1)}u_r\simeq(\ln\tilde c)'{1-\Gamma\over\Gamma(2-\Gamma)}t+
{3\over\Gamma(3-\Gamma)}{1\over r}{t^{\Gamma-1}\over\tilde c}
\left[1+\tilde c(\Gamma-1)t^{2-\Gamma}\right].
\end{equation}

Let us now examine the conditions of validity of the approximation by
determining the behaviour of the third order terms. The analysis here
is just a very particular case of the general discussion given in section 4.
Indeed we are in the case where $\mu^1_{\ 23}=0$, $u_2=u_3=0$ and $p_1=-1/3$,
$p_2=p_3=2/3$.  Therefore we know that the curvature terms do not cause
any trouble, that the condition $u_au^a <1$ implies $\Gamma>2/3$, and
finally that one must have $\Gamma>4/3$ in order to insure the validity of
equation \ref{s5} giving the energy density.  The reason for which
(\ref{s30}) fails to give the energy density when $\Gamma<4/3$ is that
the leading order in (\ref{s30}) compensate (the Kasnerian metric is a
vacuum solution), and that the subdominant term is superseded, when
$\Gamma<4/3$, by third order terms coming from the first iteration.  Let
us recover these results directly by determining the behaviour of the
third order correction.

At leading order the dominant term in the r.h.s. of (\ref{s3}) is $\bar
R\simeq-{2\over3r^2}t^{-4/3}$; as for $\bar S$ it remains negligible ($\bar
S\propto t^{\Gamma-4/3}$).  The first order scale factor being at
leading order proportional to $t^{1/3}$ we have that the third order
correction to $A$, built out of the leading part of the first order
solution is:
\begin{equation}
{}^{(1)+(3)}\left({A\over a^3}\right)
\simeq-{4\over3t}\left(1-{3\over2r^2}t^{2/3}\right)
\end{equation}
which has to be compared with ${}^{(1)}(A/ a^3)$, built with the more
accurate first order solution (\ref{s28}-\ref{s29}), that is:
\begin{equation}
{}^{(1)}\left({A\over a^3}\right)=-{4\over3t}\left[1-\tilde ct^{2-\Gamma}
+{\cal O}\left(t^{4-2\Gamma}\right)\right].
\label{s33}\end{equation}
We therefore see, in agreement with the general discussion of section 4, that
indeed the first order solution is a good approximation to a generic
solution of Einstein's equations near a spacetime singularity up to {\it
and including} terms in $t^{2-\Gamma}$ provided that $2/3>2-\Gamma$ that
is $\Gamma>4/3$.  When $\Gamma<4/3$ the metric (\ref{s28}-\ref{s29}) is
still good at leading order near the singularity but the energy density
cannot any longer be given by (\ref{s30}).

\subsection{The case  of  stiff fluid ($\Gamma=2$)}

The first order metric is again obtained by particularizing the results of
section 3 to the case of spherical symmetry. The anisotropy matrix depends on
$r$ only: ${}^{(1)}A=A(r)$; the scale factor is given by (\ref{e16}):
 ${}^{(1)}a=\bar a
u^{1/3}, u\equiv t-t_0$; and the metric (\ref{e20}) becomes:
\begin{equation}
{}^{(1)}\gamma_{rr}=\bar C_ru^{2/3-4\alpha/3}\qquad ,\qquad
{}^{(1)}\gamma_{\theta\theta}=\bar C_\theta u^{2/3+2\alpha/3}
\label{s34}\end{equation}
(with $\alpha\equiv 3A/4\bar a^3$). It depends on 4 functions: $\bar C_r,
\bar C_\theta, t_0, \alpha$. We fix the gauge by choosing $\bar C_\theta=r^2$
and $t_0=0$ so that the generic first order line element, together with the
energy density, given by (\ref{e25}) and the radial velocity derived from
(\ref{s6}) are:
\begin{equation}
{}^{(1)}ds^2=-dt^2+t^{2/3}\left[\bar C_rt^{-4\alpha/3}
+r^2t^{2\alpha/3}d\Omega^2\right]
\end{equation}
\begin{equation}
\chi{}^{(1)}\epsilon={1-\alpha^2\over3t^2}
\end{equation}
\begin{equation}
{}^{(1)}u_r={t\over1-\alpha^2}\left[\alpha\alpha'\ln
t+{3\alpha+r\alpha'\over r}\right].
\end{equation}
The solution depends on the 2 arbitrary functions:  $\bar C_r$ and
$\alpha$, and the positivity of $\epsilon$ imposes $\alpha^2<1$.

To obtain the metric at third order in the gradients we must first
evaluate the r.h.s.  of equation (\ref{s3}), that is compute $\bar R$
and $\bar S$ by means of the metric (\ref{s34}).  These are sums of
terms in $t^{{-2\over3}(1-2\alpha)}\left[ Const., \ln t, (\ln
t)^2\right]$ and in $t^{-{2\over3}(1+\alpha)}$, that we shall denote
collectively by $Q$.  Integrating (\ref{s3}) and (\ref{s4}) will then
yield ${}^{(3)}A``="A+Qt^2$ and ${}^{(3)}a``="{}^{(1)}a(1+Qt^2)$, so
that the third order metric, given by (1) will be of the form:
\begin{equation}
{}^{(1)+(3)}\gamma_{rr}={}^{(1)}\gamma_{rr}(1+\gamma_r)\qquad ;\qquad
{}^{(1)+(3)}\gamma_{\theta\theta}={}^{(1)}\gamma_{\theta\theta}
(1+\gamma_\theta)
\end{equation}
where the time dependence of $\gamma_r$ and $\gamma_\theta$ is $Qt^2$. Hence
we see that the iteration scheme is valid if $Qt^2<1$. Since $\alpha^2<1$ this
condition is satisfied near $t=0$, in agreement with the general discussion of
section 4.

The detailed calculation gives, for $\alpha>0$ and at leading order near the
singularity:
\begin{equation}
\gamma_r\simeq {9(1-2\alpha)\over r^2(7-2\alpha)(2-\alpha)^2}
t^{{2\over3}(2-\alpha)}
\end{equation}
\begin{equation}
\gamma_\theta\simeq{9(4\alpha-5)\over 2r^2(7-2\alpha)(2-\alpha)^2}
t^{{2\over3}(2-\alpha)}.
\end{equation}
(For $\alpha<0$ the dominant third order term near the singularity is in $(\ln
t)^2t^{{4\over3}(1+\alpha)}$.)
One can also check that the conditions of validity coming from the
three-velocity are also satisfied since $\Gamma=2$ and $p_a<1$.

\section{Conclusions}

In this work, we have studied the early time behaviour of inhomogeneous
spacetimes near the singularity. To do that we have used a long
wavelength iteration scheme to approximate the Einstein equations.
Our  main concern was  to test the validity of the approximation
scheme, by comparing the terms we ignored with those we kept.  The
result of this investigation is that one should be very cautious with
the use of the long wavelength approximation if one wishes to get {\it
general} results.  Indeed our analysis shows that, in the general case,
there are very severe restrictions on the range of validity of this
scheme.  The troubles arise from two origins:
\begin{itemize}
\item The curvature terms: the curvature terms, which are ignored in the
first step, blow up in general near the singularity. Going beyond
the long wavelength approximation by keeping them from the beginning
enables us to recover the oscillatory behaviour discovered by
BKL. In the case where the curl of the vector field representing the axis
of contraction (going forwards in time) is orthogonal to the vector field,
 then
the curvature terms can be ignored and the approximation scheme is valid.
\item the velocity terms: the velocity terms may also blow up.

They do not if the perfect fluid is sufficiently ``stiff'' to compensate
for the dilatation ($\Gamma>2p_a$) or if the component of the velocity
along the dilatation (time going forwards) direction vanishes.
Even if these conditions are fulfilled it is not yet enough to get a valid
energy equation, which demands that $\Gamma>1+p_a$ unless the component
$u_a$ vanishes.
\end{itemize}

In view of these results it should be interesting to reconsider the
study of BKL when the velocity terms are dominant over the curvature
terms. In this case the role of matter should become important and
one should not be able to restrict oneself to the case of vacuum.

Finally we stress the fact that the problem of the velocity terms
disappears in the case of a cosmological constant ($\Gamma=0$), and in the
case of {\it irrotational dust} where it is possible to choose a
synchronous system of coordinates for which the three-velocity of dust
is always zero.  If we impose spherical symmetry the complete scheme
works for $\Gamma>4/3$ (including $\Gamma=2$).  However the scheme works
weakly, i.e.  without the energy relation, for $\Gamma>2/3$.  The case
$\Gamma=1$ is special since one can choose a coordinate system so that
the scheme works.  The scheme also works in general for the stiff case
as soon as there is local expansion along {\bf all} the spatial
directions.

All our conclusions should apply to a gravitational collapse, instead
of a Big-Bang, by just reversing the time.

\acknowledgments
We thank A.A.  Starobinski, T.  Piran, K.  Tomita and J.  Stewart for
useful conversations.  D.L.  acknowledges support from Projet Alliance,
the Ecole polytechnique and a Golda Meir postdoctoral fellowship.  N.D.
acknowledges support from Projet Arc-en-Ciel.

\end{document}